\documentclass[twocolumn,twoside,slac_two]{revtex4}
\usepackage{graphicx}
\usepackage{fancyhdr}
\begin{document}

\title{Constraining blazars distances with combined GeV and TeV data}

\author{E.~Prandini}
\affiliation{Padova University \& INFN}

\author{M.~Mariotti}
\affiliation{Padova University \& INFN}

\author{F.~Tavecchio}
\affiliation{Osservatorio Astronomico di Brera}

\begin{abstract}

Recently, a new method to constrain the distances of blazars with unknown redshift using combined observations in the GeV and TeV regimes has been developed. The underlying assumption is that the Very High Energy (VHE, E $>$ 100\,GeV) spectrum corrected for the absorption of TeV photons by the Extragalactic Background Light (EBL) via photon-photon interaction should still be softer than the gamma-ray spectrum observed by Fermi/LAT. The constraints found are related to the real redshifts by a simple linear relation, that has been used to infer the unknown or uncertain distance of blazars. The sample is revised with the up-to-date spectra in both TeV and GeV bands and the method applied to the unknown distance blazar PKS~1424+240 detected at VHE.

\end{abstract}

\maketitle

\thispagestyle{fancy}

\section{Introduction}

\subsection{TeV Blazars}
The large majority of extragalactic TeV photons emitters
belongs to the class of blazars:
radio-loud Active Galactic Nuclei (AGN) with a 
relativistic jet closely oriented towards the Earth \cite{urry}.

The typical spectrum emitted by a blazar is 
non thermal and covers the entire electromagnetic spectrum, 
as sketched in Fig~1. It is composed by two bumps: 
at low energies the emission is synchrotron radiation 
by relativistic electrons, while at higher 
energies the origin of the radiation
is more uncertain. 
The most credited models, referred as leptonic models, involve the
inverse Compton scattering mechanism to explain the high energy emission.  
However, alternative models taking into account the presence of a hadronic 
component in the emission are not ruled out.

At GeV and TeV regimes, the photon energy flux emitted by a blazar 
is usually well approximated with a power 
law\footnote{Indeed, this is true at a reasonable distance from the 
peak maximum.} of the form 
$dN/dE = f_0 (E/E_0)^{-\Gamma}$. 

\subsection{EBL absorption}
An important effect involving VHE photons emitted by blazars
is the production of electron-positron pairs 
($\gamma \gamma \rightarrow e^{+}e^{-}$), caused by the interaction 
with the EBL \cite{stecker92}.
EBL is composed of stellar light emitted and partially 
reprocessed by dust throughout the entire history of 
cosmic evolution. 

Due to the lack of direct EBL knowledge, many models
have been elaborated in the last
years~\cite{dominguez10,franceschini08,kneiske10,stecker06}, but
the uncertainties remains quite large.

Quantitatively, the effect of the interaction of VHE photons with EBL
is an exponential attenuation of the flux 
by a factor $\tau (E,z)$, where $\tau$
is the optical depth, function of 
both photon energy and source redshift. 
This  is represented in Fig.~1, where, due to 
the absorption, the observed TeV spectrum (continuous line) differs
significantly from the emitted spectrum (dashed line).
The observed differential energy spectrum from a blazar 
is related to the emitted one according 
to $F_{\rm obs}(E)=e^{-\tau(E,z)} F_{\rm em}(E)$.

In principle it is possible to derive 
the emitted (or intrinsic) spectrum by deabsorbing
the observed spectrum. This procedure depends
on the absorption coefficient $\tau$ and the 
redshift $z$ of the source.
Vice versa, if the intrinsic source spectrum is known, 
given the absorption coefficient  $\tau$,  
the redshift $z$ can be estimated comparing 
the absorbed spectrum with the observed one. 
Here, we use the second approach.

 \begin{figure}
   \centering  
   \includegraphics[width=3.5in]{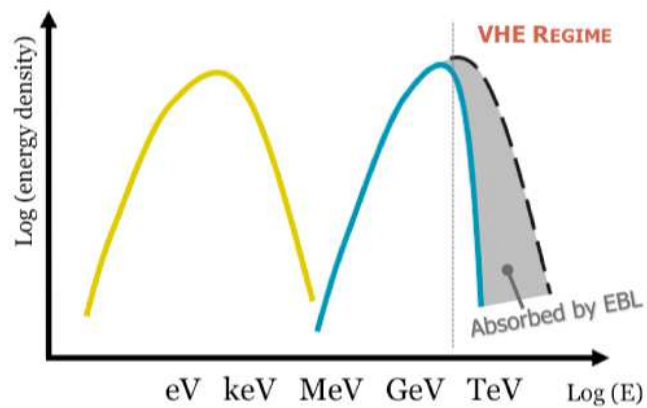}
   \caption{Simplified SED of a blazar. The observed synchrotron peak (yellow) and high energy peak (blue) are represented. The effect of EBL absorption at VHE is also represented.}
   \label{SED_blazar}
 \end{figure} 

 \begin{table*} 
       \centering
       \begin{tabular}{||l||l||c||c||c||}
         \hline
         \hline
         Source Name     & $z_{true}$  &  $\Gamma_{LAT}$ & $z^*$ & $z_{rec}$ \\ 
         \hline
        \hline
        Mkn 421          & 0.030 & 1.81 $\pm$ 0.02  &   0.07 $\pm$ 0.02  & 0.02 $\pm$ 0.05 \\
        Mkn 501          & 0.034 &  1.85 $\pm$ 0.04  &   0.08 $\pm$ 0.02 &  0.03 $\pm$ 0.05 \\
        1ES 2344$+$514   & 0.044 &  1.57 $\pm$ 0.17  &   0.19 $\pm$ 0.03 &  0.09 $\pm$ 0.05\\
        Mkn 180          & 0.045 &  1.86 $\pm$ 0.11  &   0.21 $\pm$ 0.11 &  0.11 $\pm$ 0.05\\
        1ES 1959$+$650   & 0.047 & 2.09 $\pm$ 0.05 &    0.07 $\pm$ 0.03  &  0.02 $\pm$ 0.05\\
        BL Lacertae      & 0.069 & 2.37 $\pm$ 0.04 &    0.27 $\pm$ 0.14  &  0.14 $\pm$ 0.05\\
        PKS 2005$-$489   & 0.071 &  1.90 $\pm$ 0.06 &    0.18 $\pm$ 0.03 &  0.09 $\pm$ 0.05\\
        W Comae          & 0.102 & 2.06 $\pm$ 0.04 &    0.24 $\pm$ 0.05  &  0.13 $\pm$ 0.05\\
        PKS 2155$-$304   & 0.116 & 1.91 $\pm$ 0.02 &    0.22 $\pm$ 0.01  &  0.11 $\pm$ 0.05\\
        RGB J0710$+$591  & 0.125 &  1.28 $\pm$ 0.21 &    0.21 $\pm$ 0.06 &  0.11 $\pm$ 0.05\\
        1ES 0806$+$524   & 0.138 &  2.09 $\pm$ 0.10 &    0.23 $\pm$ 0.15 &  0.12 $\pm$ 0.05\\
        H 2356$-$309     & 0.165 &  2.10 $\pm$ 0.17 &    0.16 $\pm$ 0.07 &  0.08 $\pm$ 0.05\\
        1ES 1218$+$304   & 0.182 & 1.70 $\pm$ 0.08 &    0.21 $\pm$ 0.08  &  0.11 $\pm$ 0.05\\
        1ES 1101$-$232   & 0.186 &1.36 $\pm$ 0.58 &    0.23 $\pm$ 0.11  &  0.12 $\pm$ 0.05\\
        1ES 1011$+$496   & 0.212 &  1.93 $\pm$ 0.04 &    0.60 $\pm$ 0.18  & 0.35 $\pm$ 0.05\\
        S5 0716$+$714    & 0.310$^a$ & 2.15 $\pm$ 0.03 &    0.23 $\pm$ 0.10 & 0.12 $\pm$ 0.05\\   
        PG 1553+113      & 0.400 &  1.66 $\pm$ 0.03 &    0.75 $\pm$ 0.07 &  0.45 $\pm$ 0.05\\
        3C~66A           & 0.444$^a$   & 1.92 $\pm$ 0.02 &    0.39 $\pm$ 0.05 & 0.22 $\pm$ 0.05 \\
        \hline
        \hline
      \end{tabular}
      \caption{-- List of TeV blazars used in this study. Source name (first column), $\Gamma_{LAT}$ reported in the first year catalogue (second column), $z^*$ and $z_{rec}$ values obtained with the Franceschini EBL model (fourth and fifth columns). $^a$: uncertain. \label{table_values}}
  \end{table*}

\section{The Method}

In a recent paper, we have proposed a method to derive an estimate on 
the distance of a blazar \cite{prandini10}.
The method is based on the comparison between the spectral index at GeV energies
as measured by LAT after 5.5 months of data taking,  basically 
unaffected by the cosmological absorption, 
and the TeV spectrum corrected for the absorption. 
In that work, it is  shown that according to present observations,
the spectral slope measured by LAT, 
$\Gamma_{LAT}$, in the energy range 0.2\,--\,300\,GeV, 
can be considered as a limiting slope for the 
emitted spectrum at TeV energies (i.e. corrected for EBL absorption). 
This maximum hardness hypothesis was successfully 
tested on a sample of 14 well--known distance sources.
Consequently, the redshift, $z^*$, at which 
the deabsorbed TeV slope equals $\Gamma_{LAT}$, 
can be used as an upper limit on the source distance.

An empirical relation between the
upper limit, $z^*$, and the true redshift of a blazar was then found. 
A simple linear relation fits well the $z^*$\,--\,$z_{true}$ distribution, 
for three different EBL models. 
The relation is  associated to the
linear expression, found in \cite{stecker10}, for the steepening of the observed 
TeV slopes due to EBL absorption.
Hence, $z^*$ and $z_{true}$ 
are related by a linear function of the form $z^*=A+Bz_{true}$.
This relation can be used to 
give an estimate on the source distance.

In this paper we present an update of that work, based on a more
recent LAT catalogue \cite{abdo10}. We test the validity of the
maximum hardness hypothesis and that of the linear relation
between $z^*$ and $z_{true}$.
A cosmological scenario with $h=0.72$, $\Omega_M=0.3$ 
and $\Omega_\Lambda=0.7$ is assumed. 

  \begin{figure*}
    \centering  
    \includegraphics[width=5.2in]{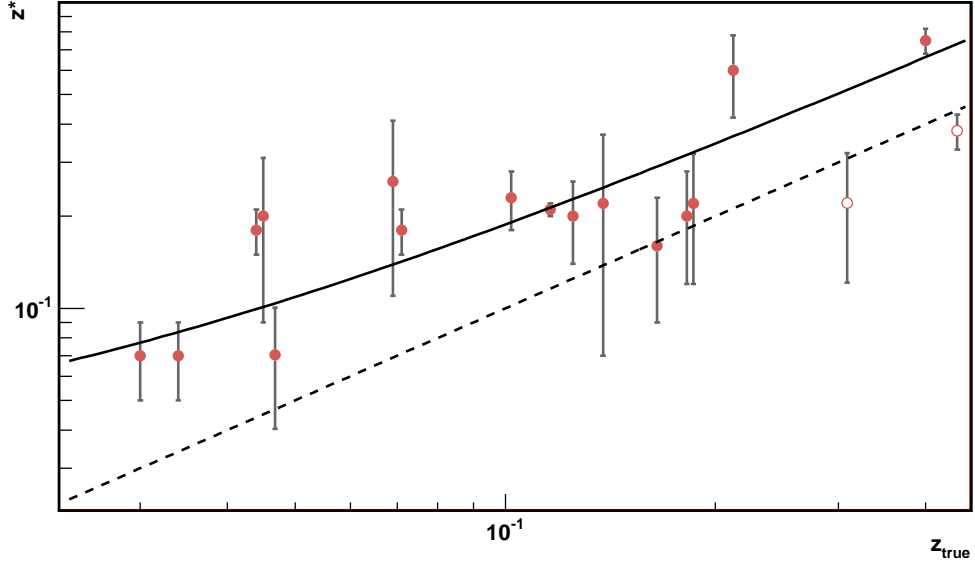}
    \caption{ $z^*$ versus true redshift derived with the 
      procedure described in the text. 
      The open symbols represent the two sources with
      uncertain redshift, 3C~66A and S5~0716+714, not used in 
      the fit. The dashed line is the bisector, 
      while the continuous line is the 
      linear fit to the data.}
    \label{correlationplot}
  \end{figure*} 

\section{Analysis and results}
The sample presented in this study
is composed by all the extragalactic TeV 
emitters located at redshift larger than $z\,=\,0.01$ and 
detected by LAT after the first year of data taking \cite{abdo10}.
In total, there are 16 sources with well known redshift and two
additional sources with uncertain redshift, namely 3C~66A and S5~0716+714. 
In the first column of Table~\ref{table_values} we list the 
sources used in the study. The second column represents
the slope measured by LAT after the first year of data taking, in the 
energy range 0.1\,--100\,GeV.
Three new sources are added to the sample considered in the 
original study, namely: 
RGB~J0710$+$591, H~2356$-$309 and 1ES~1101$-$232, 
located at redshifts 0.125, 0.165 and 0.186, respectively. 
The last two sources were not detected by LAT in the first 5.5 months, 
while the spectrum of RGB~J0710$+$591 has only
recently been published by the VERITAS collaboration \cite{acciari10b}. 
With respect to the 5.5 months catalogue, the new LAT determination of the
spectral slopes is characterized by smaller errors, 
due to the increased statistics. 

With this enlarged  data set, we estimate the quantity 
$z^*$, redshift at which the deabsorbed
TeV spectrum exhibits the same slope measured by LAT at lower energies. 
We adopt the energy density EBL model \cite{franceschini08}, 
hereafter Franceschini model\footnote{The absorption values
used here are taken directly from the WEB site http://www.astro.unipd.it/background,
and differ slightly  from those used in \cite{prandini10}, where an 
extrapolation method was used.}. 
The $z^*$ values obtained are listed in the fourth
column of Table~\ref{table_values}.

Figure~\ref{correlationplot} represents the distribution $z_{true}~-~z^*$
obtained. All the $z^*$ values distribute above or on the bisector.
This confirms that  $z^*$ can be considered 
an upper limit on the source redshift, hence the 
maximum hardness hypothesis is confirmed also in this study.
The linear curve drawn represents the fit of the data.
The linear trend of the distribution
is less evident here than in the previous study. 
The probability of the fit is, in fact, 
$\sim$6\%, well below the previous value (58\%).
The reason for this behaviour can be related 
to the new sources introduced, but also to the 
new LAT determination of the slopes. 
In order to investigate such alternatives, we have fitted the 
distribution excluding the
three new sources. The new fit  returns a probability of 
9\%, close to the value obtained with the full sample.
This result suggests that the low probability found is mainly due to the
smaller error bars characterizing the determination of the new slopes in 
the GeV band with respect to previous estimates.
The parameters obtained are listed in Table~\ref{table:fit_parameters}. 

\begin{table}
  \centering
  \begin{tabular}{ ||c|| c||}
  \hline
  \hline
     $A$ & $B$    \\
  \hline
  \hline
     0.036 $\pm$ 0.014  & 1.60 $\pm$ 0.14  \\
   \hline
  \hline
\end{tabular}
\caption{ -- Parameters of the linear fitting curve ($z^*$~=~$A+Bz_{true}$). \label{table:fit_parameters}}
\end{table}

  \begin{figure}
    \centering
    \includegraphics[width=0.48\textwidth]{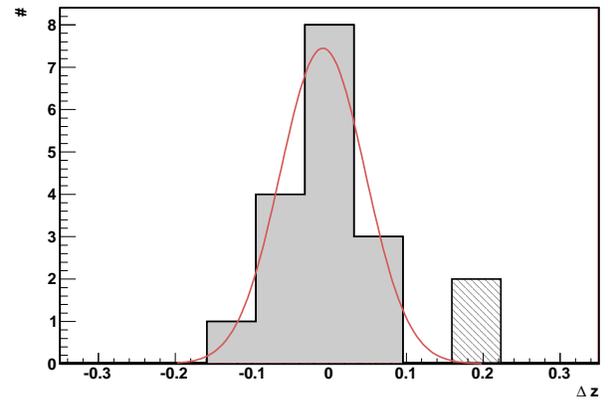}
     \caption{Dispersion $\Delta z$ ($z_{true} - z_{rec}$).
       The shaded area represents the two sources with 
      uncertain redshift (S5~0716+714 and 3C~66A), not used in the Gaussian fit.}
    \label{fig:dispersion_fradom_fermi1y}
  \end{figure}

\begin{figure*}
     \centering
     \includegraphics[width=0.6\textwidth]{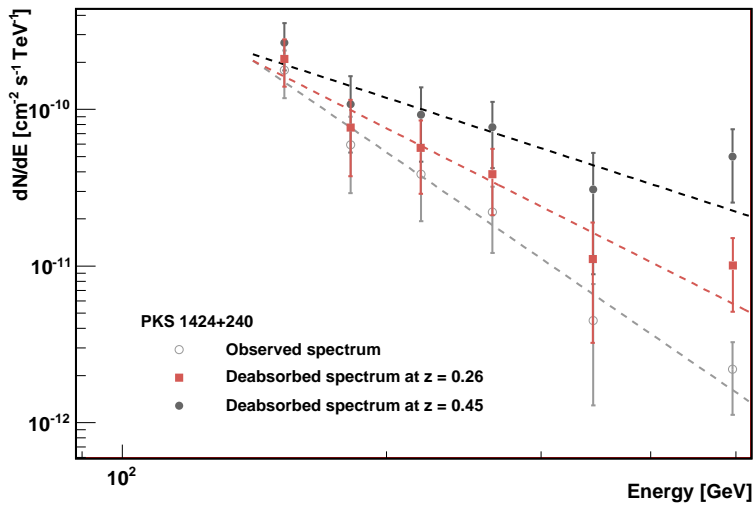}
     \caption{Measured (open circles) and deabsorbed (filled circles and squares) spectra of PKS~1424+240, assuming a redshift $z$~=~0.45 and  $z$~=~0.26, respectively.}
     \label{1424_plot}
   \end{figure*}

Following the first study, 
we investigate the distribution $\Delta z$, difference between
the values $z_{rec}$, listed in the last column of Table~\ref{table_values}, obtained
by inverting the linear formula $z_{rec}$~=~($z^*-A)/B$, and the true redshifts, $z_{true}$. 
The histogram obtained, Figure~\ref{fig:dispersion_fradom_fermi1y},
is well fitted by a Gaussian of $\sigma~=~0.05$, 
which can be assumed as the error on the 
reconstructed redshift, $z_{rec}$, estimated with this method. 
In the histogram, the two sources with uncertain redshift, not used
for the Gaussian fit, lie outside the expected interval. 
This result confirms that the behaviour of S5~0716+714 and 3C~66A
is different from that found for other sources and suggests that or 
these sources are peculiar, or their redshift is incorrect.

In conclusion, we can say that with an enlarged data set 
the results previously found are confirmed. However, the 
linearity of the $z^*$--$z_{true}$ relation has a smaller 
probability, due to the reduced 
errors of the new $\Gamma_{LAT}$ determinations.

\section{The redshift of PKS~1424+240}
As a final application, we use our method on
PKS~1424+240, a blazar of unknown redshift
recently observed in the VHE regime by VERITAS \cite{acciari10}. 
The slope  measured by {\it Fermi}/LAT in the energy 
range 0.1\,--\,100\,GeV is $1.83\pm0.03$. 

The corresponding $z^*$ at which the slope of the deabsorbed TeV spectrum 
becomes equal to it is $0.45~\pm~ 0.15$, filled circles in Fig.~\ref{1424_plot}.
This result is in agreement with the value of $0.5\pm0.1$, 
reported in \cite{acciari10}, calculated by applying the same procedure 
but using only simultaneous LAT data.

Our estimate on the most probable distance for PKS~1424+240
is 0.26\,$\pm$\,0.05, where the error is the $\sigma$ of the Gaussian 
fitting the $\Delta$z distribution. 
The deabsorbed spectrum of PKS~1424+240 assuming this distance
is drawn in Fig~\ref{1424_plot}, filled squares.

\end{document}